\documentclass{article}

\usepackage{PRIMEarxiv}

\usepackage[utf8]{inputenc} 
\usepackage[T1]{fontenc}    
\usepackage{hyperref}       
\usepackage{url}            
\usepackage{booktabs}       
\usepackage{amsfonts}       
\usepackage{nicefrac}       
\usepackage{microtype}      
\usepackage{lipsum}
\usepackage{fancyhdr}       
\usepackage{graphicx}       
\graphicspath{{media/}}     

\usepackage{amssymb}
\usepackage{csquotes}
\usepackage{textgreek}
\usepackage{float}
\usepackage{siunitx}
\usepackage{stmaryrd}

\usepackage{physics}
\usepackage{multirow}

\usepackage{caption}
\usepackage{subcaption}
\usepackage{xcolor}

\usepackage{booktabs}

\usepackage{empheq, etoolbox}
\title{Vertex and front-tracking methods for the modeling of microstructure evolution at the solid state: a brief review}

\author{
  Marc Bernacki \\
  Mines Paris, PSL University\\
  Centre for material forming (CEMEF), UMR CNRS\\
  06904 Sophia Antipolis, France\\
  \texttt{marc.bernacki@minesparis.psl.eu} \\
}

\begin{document}
\maketitle

\begin{abstract}
In mesoscopic scale microstructure evolution modeling, two primary numerical frameworks are used: Front-Capturing (FC) and Front-Tracking (FT) ones. FC models, like phase-field or level-set methods, indirectly define interfaces by tracking field variable changes. On the contrary, FT models explicitly define interfaces using interconnected segments or surfaces. In historical FT methodologies, Vertex models were first developed and consider the description of the evolution of polygonal structures in terms of the motion of points where multiple boundaries meet.  Globally, FT-type approaches, often associated with Lagrangian movement, enhance spatial resolution in 3D surfacic and 2D lineic problems using techniques derived from finite element meshing and remeshing algorithms. These efficient approaches, by nature, are well adapted to physical mechanisms correlated to interface properties and geometries.  They also face challenges in managing complex topological events, especially in 3D. However, recent advances highlight their potential in computational efficiency and analysis of mobility and energy properties, with possible applications in intragranular phenomena.
\end{abstract}


\section{Introduction}
When dealing with the modeling of microstructural evolutions at the mesoscopic scale, there are typically two main families of methods to describe complex moving or interfaces network: Front-Capturing (FC) and Front-Tracking (FT) ones. This terminology is ultimately common to all themes where the monitoring of a front or a network of interfaces is central \cite{DeSousa2004}. FC models, like phase-field/ multiphase field \cite{Steinbach1996,Moelans2008,KrillIII2002} or level-set \cite{Osher1988, Dervieux1980,Merriman1994,Zhao1996,Bernacki2008,Hallberg2019,Bernacki2024} frameworks implicitly (indirectly) define interfaces by capturing the change of state of a field variable. In contrast, FT models explicitly (directly) define interfaces using a set of interconnected segments (in 2D) or surfaces (in 3D). \\

By this definition, Vertex models \cite{Fullman1952} are FT models, as the vertices are connected between them to form the segments that define the boundaries between domains. It will be illustrated that the boundary between the definition of Vertex and front-tracking methods, and even the definition of front-tracking methods themselves, have evolved over time compared with the first articles on the subject, and that can sometimes be rather vague. \\

Typically, approaches such as Monte Carlo \cite{rollett1989computer,Rollett2004a,radhakrishnan1998monte,radhakrishnan1998modeling,zhang2012calibrated,wang2014monte,sieradzki2013perceptive,Villaret2020} or Cellular Automaton \cite{Raabe2002,Janssens2010,golab2014,madej2023computationally,sieradzki2013perceptive} can be considered from the point of view of recent authors as FT-type approaches since, in these frameworks, the grain boundaries can be explicitly described by a collection of edges of pixels/voxels used to discretize the computational domain. 
Historically, it was rather the fact that the physics of the mechanism was implemented solely through the migration of the interface that distinguished the front-tracking approaches from the preexisting approaches (MC and CA) \cite{Frost1996}. \\

The notion of an FT approach is thus today often associated with a Lagrangian movement of interfaces to describe the movement of modeled interfaces networks and for which the kinetics are derived from geometrical properties which are only dependent of the interfaces. This can therefore be done mainly by using techniques based on FE meshing/remeshing techniques or by approaches limiting the interfaces discretization and thus the resolution cost (as Vertex approaches). A common interest to these approaches is to gain a degree of spatial resolution compared to other existing approaches by considering surface geometries in 3D and line ones in 2D. The flip side of the coin is that these methods are generally not suitable for modeling intragranular mechanisms. Thus, the most often modeled mechanism concerns grain growth problems without heterogeneous plastic stored energy field in the core of the grains and where only the surface energy of the interfaces network needs to be minimized. The difficulty of managing topological operations at multiple junctions is also a complex algorithmic subject, especially in 3D. Again, this explains the limited number of studies dedicated to the use of these approaches in 3D for a large number of interfaces. However, recent work illustrates more than ever the interest of these methods in terms of computational time optimization, precise discussion of the mobility and energy properties at multiple junctions, and their possible extension for the consideration of intragranular phenomena. These different elements will be presented below with regard to the state-of-the-art.

\section{Vertex frameworks}

\begin{figure}[ht!]
  \centering
  \includegraphics[scale=0.3]{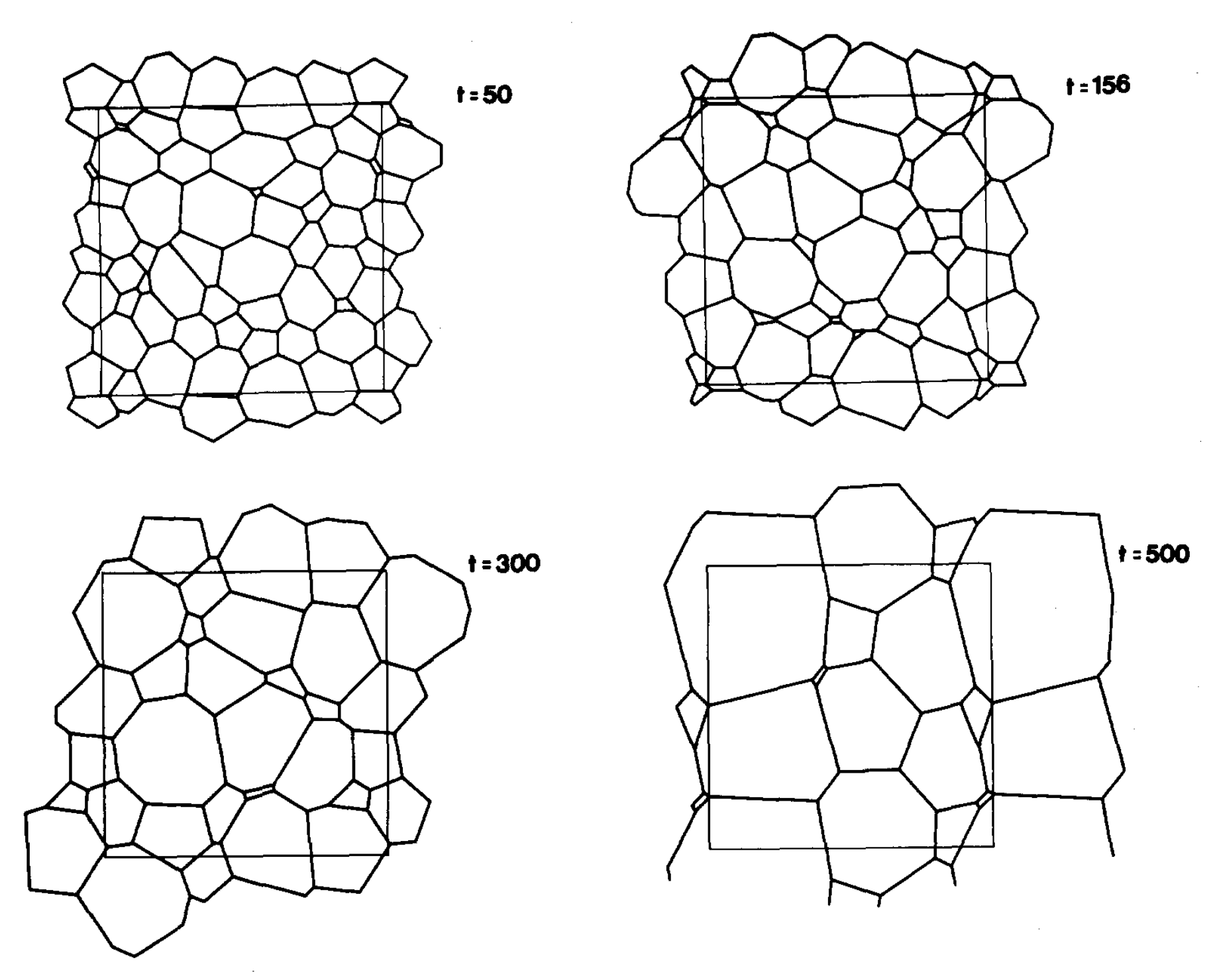}
  \caption{Successive stages in the growth of a 2D grain structure with the number of steps (proportional to the physical time). Simulation from \cite{Soares1985}.}\label{fig:fig1}
\end{figure}

\begin{figure}[!ht]
	\centering
	\includegraphics[width=0.8\textwidth]{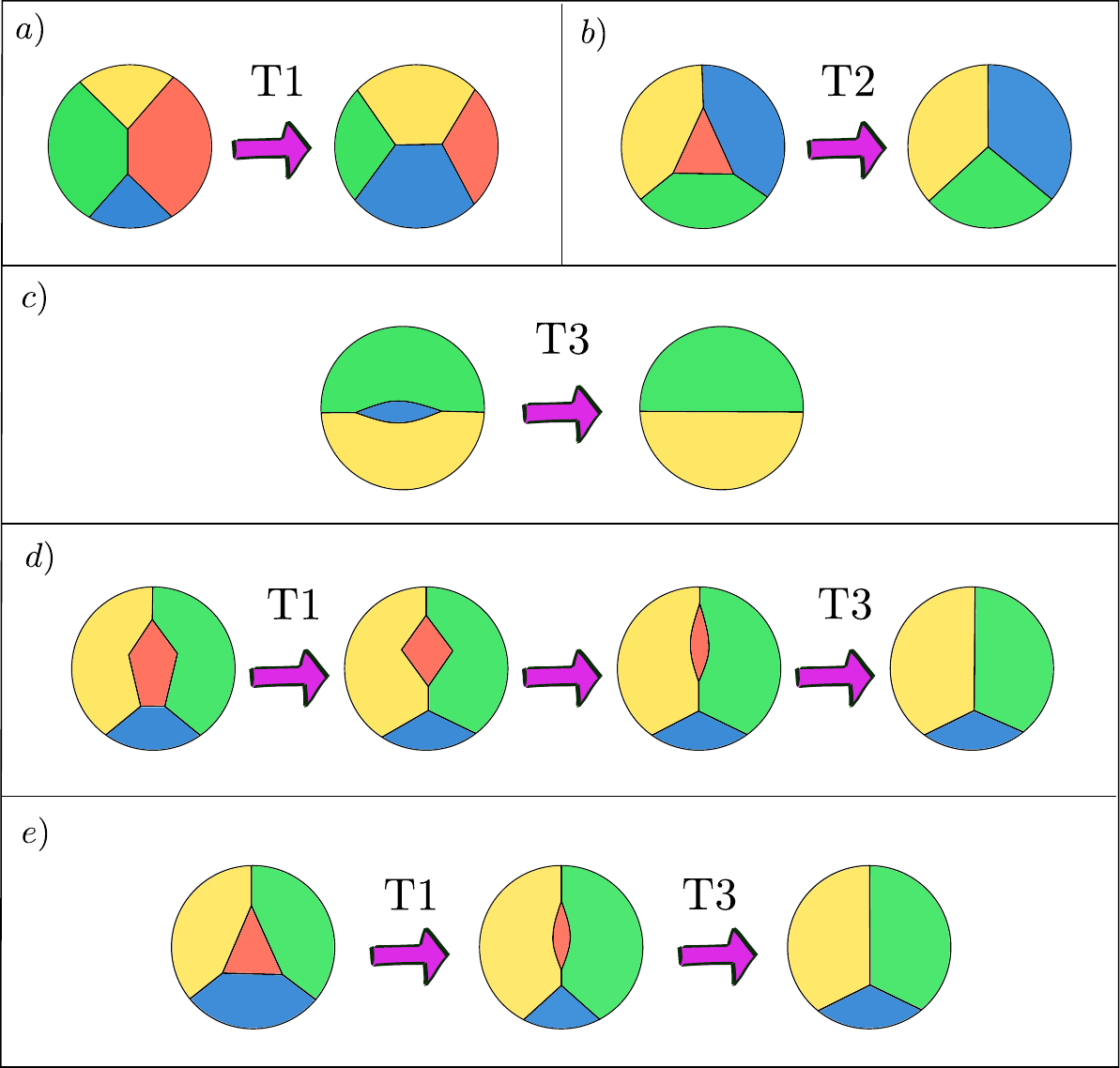}
	\caption{\label{fig:TopoOperationsWeygand} 2D Topological transformations: a) recombination (T1), b) annihilation of small three-sided grain (T2), c) elimination of a grain with two triple points (T3), d) an example of the occurrence of a T3 transformation and e) T2 transformation shown as equivalent to a sequence of T1 and T3 transformations. Source: From \cite{FlorezPhD2020} as inspired from \cite{Weygand1998First}.}
\end{figure}

In the early 1950s, John von Neumann introduced a kinetic law for two-dimensional soap froths, establishing a connection between the growth rate of individual bubbles'areas and their number of neighbors \cite{von1952metal}. This concept was later expanded by Mullins \cite{mullins1956} to encompass two-dimensional grain growth structures characterized by isotropic boundary energies. Known today as the von Neumann–Mullins law, this equation was further extended to three dimensions fifty years later by MacPherson and Srolovitz \cite{macpherson2007neumann}.
Interestingly, the appearance of the first Vertex framework in the context of 2D grain growth has followed the same logical way as the model presented by Soares et al. \cite{Soares1985} is based on the model of \cite{Weaire1983} that was applied to the evolution of soap froths. These models consider that the evolution of an interface network can be simulated from the movement of its triple junctions (multiple junctions with three connections), a proposition first suggested in \cite{Fullman1952} (but with arbitrary hypothesis concerning the velocity definition). This choice allows to avoid the computation of curvature by the use of straight lines as grain boundaries. The motion equation of the triple junctions used by Soares et al. was derived from a thermodynamics study (unlike the one proposed in \cite{Fullman1952}) and can be summarized as:

\begin{equation}
\label{Eq:SoaresVelocityTJ}
\vec{v}=\mu \sum \vec{\varepsilon}_i,
 \centering
\end{equation}

where $\vec{v}$ and $\mu$ are, respectively, the velocity and mobility of a vertex and $\vec{\varepsilon}_i$ denotes each of the line tensions acting at a vertex. Figure \ref{fig:fig1} illustrates a 2D grain growth simulation performed by Soares et al. for few grains thanks to this method. In this model, the mobility was assumed homogeneous and vertices at all times only have 3 possible connections, avoiding higher order multiple junctions configurations by means of unitary topological changes (only the  transformations \textbf{T1} and \textbf{T2} of Fig.~\ref{fig:TopoOperationsWeygand} were considered). Moreover, $\mu$ and $|{\varepsilon_i}|=\varepsilon$ were assumed homogeneous in space and constant in time; therefore, the model was tested only under isotropic conditions. In these early works, the absence of actual curvature calculations, and thus the manipulation of exclusively straight interfaces, was crucial in proposing an efficient method compared to the computational means available at that time. Interestingly, Soares et al. exhibited in this article a very good agreement with the Burke and Turnbull law \cite{BT1952} seemingly without being aware of it.\medbreak

Few years later, by keeping the idea of grain interfaces described by straight lines between evolving multiple junctions (the vertices), i.e., by the fact that the grain boundary network is defined by the positions $\vec{r}$ of the vertices and their velocities $\vec{v}$, a more general model was proposed in \cite{Kawasaki1989}. Here, the authors proposed to associate the polycrystalline structure with a potential term $\nu\left(\vec{r}\right)$ that describes the total surface energy and a dissipative term $R\left(\vec{r},\vec{v}\right)$ that describes the dissipation due to the movement of the GB network. These two terms can be summarized as :
\begin{equation}\label{eq:nu}
\nu\left(\vec{r}\right)=\int_{\Gamma}\gamma \, d\Gamma,
\end{equation}

\begin{equation}\label{eq:R}
R\left(\vec{r},\vec{v}\right)=\frac{1}{2}\int_{\Gamma}\frac{\lVert{\vec{v}\rVert}^2}{\mu} \, d\Gamma,
\end{equation}
where $\gamma$ corresponds to the grain boundary energy and $\Gamma=\partial G$ depicts here the union of the GBs, i.e. the GB network. In this scenario, the equations of motion are obtained by deriving them from the Lagrange function that includes a dissipative term but does not include a term for kinetic energy:

\begin{equation}
\label{Eq:KawasakiDissipation}
\dfrac{\partial R}{\partial\vec{v}} = - \dfrac{\partial \nu}{\partial\vec{r}}.
 \centering
\end{equation}

The Vertex method finally consists of transform Eqs \ref{eq:nu}, \ref{eq:R} and \ref{Eq:KawasakiDissipation} in a discretized way by considering the vertices and the egdes between vertices (i.e. the GB) as the geometric support of the discretization. First of all, by considering, $N_v$ the total number of vertices, $N_i$ the number of neighboring vertices to the vertice $i$ $\forall i\in\llbracket 1,N_v\rrbracket$, $\gamma_{ij}$ the GB energy in the $ij$ segment, and $\vec{r}_{ij}=\vec{r}_{i}-\vec{r}_{j}$:
\begin{equation}\label{eq:nud}
\nu\left(\vec{r}\right) =\frac{1}{2}\sum_{i=1}^{N_v}\sum_{j=1}^{N_i}\gamma_{ij}\lVert\vec{r}_{ij}\rVert.
\end{equation}

Now, considering the fact that the velocity at a point $k$ of the segment $ij$ whose position is defined by,
$$
\vec{r}=\underbrace{\frac{\lVert\vec{r}_{kj}\rVert}{\lVert\vec{r}_{ij}\rVert}}_{\xi}\vec{r}_i + \underbrace{\frac{\lVert\vec{r}_{ik}\rVert}{\lVert\vec{r}_{ij}\rVert}}_{1-\xi}\vec{r}_j,
$$
is $\vec{v}\left(k\right)=\left((\xi\vec{v}_i+\left(1-\xi)\vec{v}_{j}\right)\cdot\vec{n}_{ij}\right)\vec{n}_{ij}$ with $\vec{n}_{ij}$ a unitary orthogonal vector to $\vec{r}_{ij}$; we can operate the calculation of $R\left(\vec{r},\vec{v}\right)$:
\begin{align}
\label{eq:eqRd1}R\left(\vec{r},\vec{v}\right) & =\frac{1}{4}\sum_{i=1}^{N_v}\sum_{j=1}^{N_i}\frac{1}{\mu_{ij}}\int_{0}^{1}\left(\xi\vec{v}_i\cdot\vec{n}_{ij}+\left(1-\xi\right)\vec{v}_j\cdot\vec{n}_{ij}\right)^{2}\lVert\vec{r}_{ij}\rVert\, d\xi,\\
\label{eq:eqRd2}R\left(\vec{r},\vec{v}\right) & =\sum_{i=1}^{N_v}\sum_{j=1}^{N_i}\frac{\lVert\vec{r}_{ij}\rVert}{12\mu_{ij}}\left(\left(\vec{v}_i\cdot\vec{n}_{ij}\right)^2+\left(\vec{v}_j\cdot\vec{n}_{ij}\right)^2+\left(\vec{v}_i\cdot\vec{n}_{ij}\right)\left(\vec{v}_j\cdot\vec{n}_{ij}\right)\right).
\end{align}
Thus, the $N_v$ equations of movement are obtained by verifying Eq.\ref{Eq:KawasakiDissipation} in each vertices and by using Eq.\ref{eq:nud} and Eq.\ref{eq:eqRd2}:
\begin{align}
\label{Eq:KawasakiDissipation2}\dfrac{\partial R}{\partial\vec{v}_i} & = - \dfrac{\partial \nu}{\partial\vec{r_i}},\\
\label{eq:nu2}\dfrac{\partial \nu}{\partial\vec{r}_i} & = \frac{1}{2}\sum_{j=1}^{N_i}\gamma_{ij}\frac{\vec{r}_{ij}}{\lVert\vec{r}_{ij}\rVert},\\
\label{eq:eqRd3}\dfrac{\partial R}{\partial\vec{v}_i} & =\sum_{j=1}^{N_i}\frac{\lVert\vec{r}_{ij}\rVert}{12\mu_{ij}}\left(2\left(\vec{v}_i\cdot\vec{n}_{ij}\right)\vec{n}_{ij}+\left(\vec{v}_j\cdot\vec{n}_{ij}\right)\vec{n}_{ij}\right),
\end{align}
By using the fact that $\vec{n} \otimes \vec{n}\vec{u}=\left(\vec{u}\cdot\vec{n}\right)\vec{u}$, the last equation can be rewritten as :
\begin{equation}\label{eq:eqR4}
\dfrac{\partial R}{\partial\vec{v}_i} =\sum_{j=1}^{N_i}\frac{\lVert\vec{r}_{ij}\rVert}{6\mu_{ij}}\vec{n}_{ij} \otimes \vec{n}_{ij}\left(\vec{v}_i + \frac{1}{2}\vec{v}_j\right).
\end{equation}
Finally, by combining Eq.\ref{Eq:KawasakiDissipation2}, Eq.\ref{eq:eqR4}, and Eq.\ref{eq:nu2}, we obtain the well-known Kawazaki et al. equation:
\begin{equation}\label{Eq:KawasakiImplicit}
\dfrac{1}{3}\sum_{j=1}^{N_i}\frac{\lVert\vec{r}_{ij}\rVert}{\mu_{ij}}\vec{n}_{ij} \otimes \vec{n}_{ij}\left(\vec{v}_i + \frac{1}{2}\vec{v}_j\right)
 = -  \sum_{j=1}^{N_i}\gamma_{ij}\frac{\vec{r}_{ij}}{\lVert\vec{r}_{ij}\rVert},
\end{equation}
i.e.
\begin{equation}\label{Eq:KawasakiImplicit2}
\mathcal{D}_{i}\vec{v}_i=\vec{f}_i - \frac{1}{2}\sum_{j=1}^{N_i}\mathcal{D}_{ij}\vec{v}_j,\text{ with,}
\end{equation}
$$
\mathcal{D}_{ij}=\frac{\lVert\vec{r}_{ij}\rVert}{3\mu_{ij}}\vec{n}_{ij} \otimes \vec{n}_{ij},\ \mathcal{D}_{i}=\sum_{j=1}^{N_i}\mathcal{D}_{ij}\text{, and }\vec{f}_i =-\sum_{j=1}^{N_i}\gamma_{ij}\frac{\vec{r}_{ij}}{\lVert\vec{r}_{ij}\rVert}.
$$
As highlighted by Kawazaki et al. thirty-five years ago \cite{Kawasaki1989}, this equation is not very comfortable to handle since the determination of $\vec{v}_i$ cannot be calculated directly from the positions of $\vec{r}_{i}$ and the neighboring vertices. Thus, they proposed to deal with simplified formulations known as Model I and Model II. In Model II, the anisotropy is neglected and an average is done over all directions of $\vec{n}_{ij}$ and $\vec{v}_j$ leading to simplified dissipative term, $D_i$, and kinetic equations:

\begin{equation}\label{eq:eqRd2II1}
R^{II}\left(\vec{r},\vec{v}\right)  =\sum_{i=1}^{N_v}\sum_{j=1}^{N_i}\frac{\lVert\vec{r}_{ij}\rVert}{12\mu}\lVert\vec{v}_{i}\rVert ^{2},\ \mathcal{D}_{i}  =\frac{1}{6\mu}\sum_{j=1}^{N_i}\lVert\vec{r}_{ij}\rVert,
\end{equation}
\begin{equation}\label{eq:eqRd2II2}
\mathcal{D}_{i}\vec{v}_i  =\vec{f}_i\text{, i.e. } \vec{v}_i = \frac{6\mu}{\sum_{j=1}^{N_i}\lVert\vec{r}_{ij}\rVert}\vec{f}_i .
\end{equation}

Moreover, in Model-I, the term $\sum_{j=1}^{N_i}\lVert\vec{r}_{ij}\rVert$ is approximated as $3r\left(t\right)$ with $r\left(t\right)$ the mean value of the edge length, leading to 
\begin{equation}\label{eq:eqRd2I1}
\vec{v}_i = \frac{2\mu}{r\left(t\right)}\vec{f}_i .
\end{equation}

The Model-I and Model-II were tested by Kawazaki et al. in \cite{Kawasaki1989} by considering the T1 and T2 transformations introduced earlier in \cite{Weaire1983} (see the top of Fig.~\ref{fig:TopoOperationsWeygand}).

\section{From Vertex to front-tracking then to enriched Vertex frameworks}

\begin{figure}[!ht]
	\centering
	\includegraphics[width=0.8\textwidth]{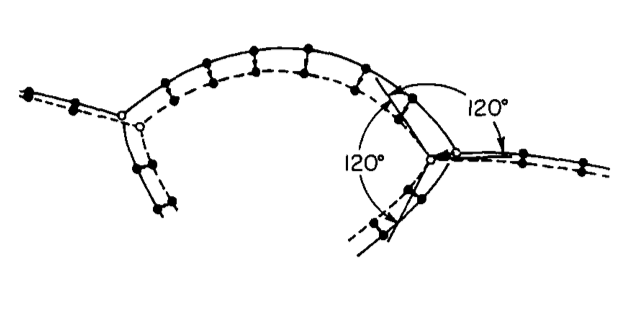}
	\caption{\label{fig:frost1988} Schematic illustration of simulation techniques for modeling boundary migration and triple point motion. Source: \cite{Frost1988}}
\end{figure}

It is interesting to note that almost simultaneously with the first Vertex articles for grain growth modeling \cite{Soares1985,Weaire1983}, another point of view, based on an enrichment of the description of interfaces, was proposed. The viewpoint being ultimately radically different: modeling the physical equation describing the kinetics of grain boundaries by discretizing the GB, used time-explicit Euler scheme to update the node positions at each time step, while imposing at each time the position of the triple junctions to respect a 120° equilibrium. This idea was first introduced by Frost et al. in 1988 \cite{Frost1988} as a more precise alternative to the Vertex or Monte Carlo approaches. Following the Von Neumann-Mullins kinetic law for grain growth, Frost and coworkers considered $\vec{v}=-\tilde{\mu}\kappa\vec{n}$ to evaluate the GB migration rate with $\kappa$ the local curvature and $\tilde{\mu}$ a proportionality constant that would today be called reduced mobility, that is to say the product of GB mobility and energy. This methodology is illustrated in Fig.\ref{fig:frost1988}.\\

\begin{figure}[!ht]
	\centering
	\includegraphics[width=0.95\textwidth]{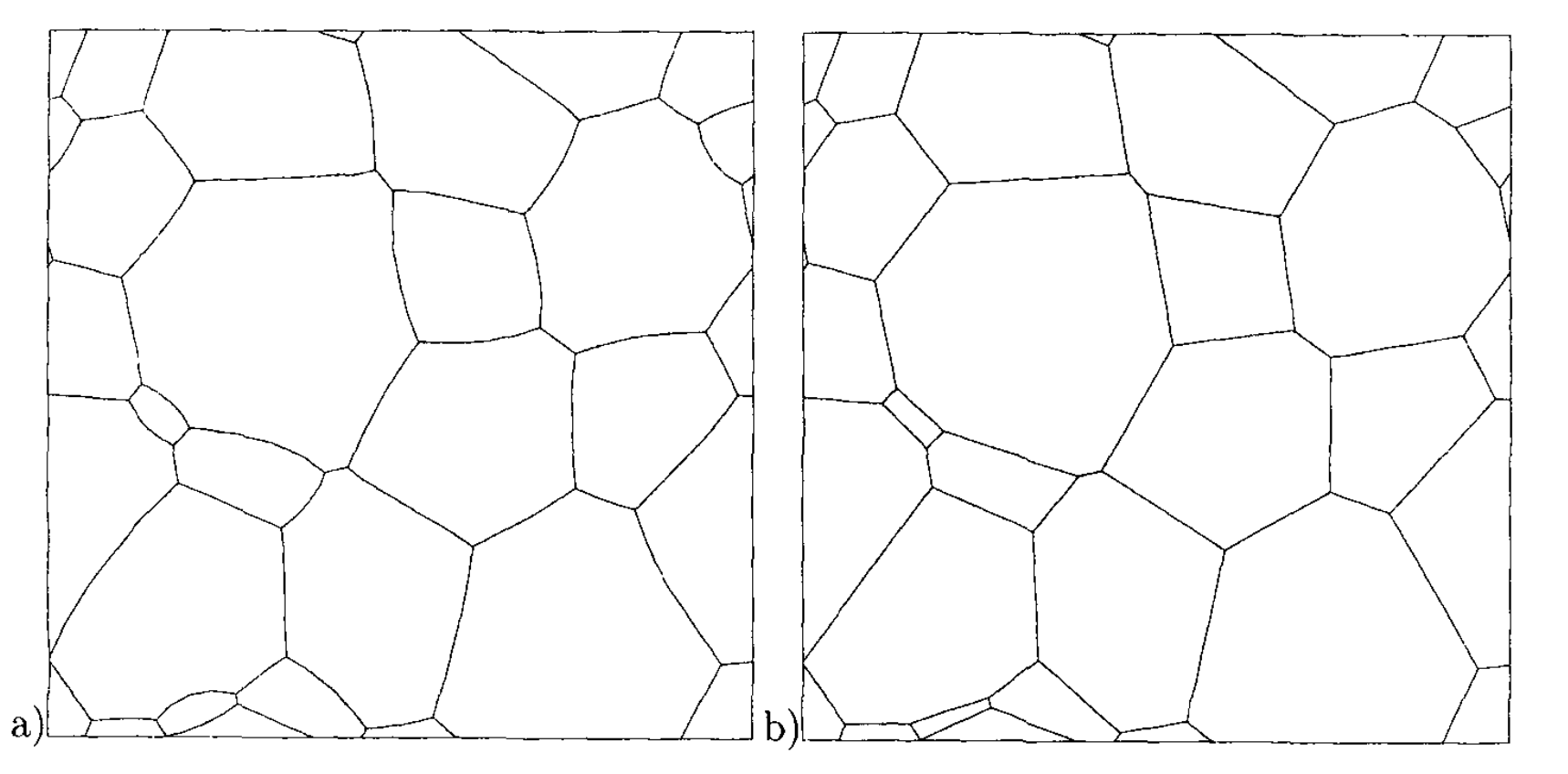}
	\caption{\label{fig:weygand1998} Grain structure obtained by Weygand et al. by considering virtual vertices alongs GBs (a) and without virtual vertices (b). Source: \cite{Weygand1998First}}
\end{figure}

Following these works, one can estimate that some confusion has arisen in the state-of-the-art between what differentiates a Vertex approach from a front-tracking type approach in general: the fact of subdividing grain boundaries? The explicit consideration of curvature in simulations? The question of verifying a kinetic equation or considering an approach with dissipative and potential terms? In the seminal work of Frost \cite{Frost1988}, the first point is highlighted as essential in order to propose a more realistic description of GB, that is, curved. The rest of the story, however, is less clear-cut. Indeed, very quickly, new approaches based on the Kawasaki et al. principle emerged while trying to better take into account curvature \cite{Marsh1995}. Finally, the works of Weygand et al. \cite{Weygand1998First} extended the concept of Kawasaki et al., by implementing a Vertex model capable of modeling curved grain boundaries. The model used \emph{real} and \emph{virtual} vertices, where virtual vertices were used to discretize the grain boundaries between triple junctions (real vertices). Weygand et al. also directly used Eq.~\ref{Eq:KawasakiImplicit2} in order to compute the velocity values of all vertices (both real and virtual) instead of the Model-II approximation Eq.~\ref{eq:eqRd2II1} used by Kawasaki et al. in their study \cite{Kawasaki1989}. An illustration of the seminal works of Weygand et al. \cite{Weygand1998First} is shown in Fig.\ref{fig:weygand1998}. Furthermore, the authors also introduce a topological operation inherent in the use of virtual vertices, corresponding to the topological operation \textbf{T3} given in Fig.~\ref{fig:TopoOperationsWeygand}.\\

Additionally, in \cite{Weygand1998First} the notion of microstepping was applied for the first time. Micro-stepping is necessary given the unstable nature of some vertices that disrupt the smoothness of grain boundaries (vertices that are too off-positioned regarding the tendency of their contiguous vertices) and which may appear after a topological change or another numerical phenomenon. Micro-stepping was then implemented using the following reasoning: no segment should shrink nor extend more than a fraction $f=dl/l$ of its length $l$ (where the reported value for $f$ was 0.5), leading to a maximum time step allowed per vertex. Finally, the movement of vertex $i$ was performed using a time-explicit Euler scheme both for the hole steps and for the microsteps. Using this model, Weygand et al. found that multiple junctions evolve to fulfill their equilibrium angles.\\

Today, ambiguity sometimes persists, as some articles refer to the use of the Kawasaki principle to discuss the Vertex approach. There are numerous Vertex papers discussing kinetic approaches based on curvature to describe interfaces, themselves discretized. From the author perspective, these approaches align more closely with front-tracking methodologies, as originally conceptualized in \cite{Frost1988}, rather than the original Vertex methods.
\section{Enrichment of the Vertex approach}
\begin{figure}[!ht]
	\centering
	\includegraphics[width=0.8\textwidth]{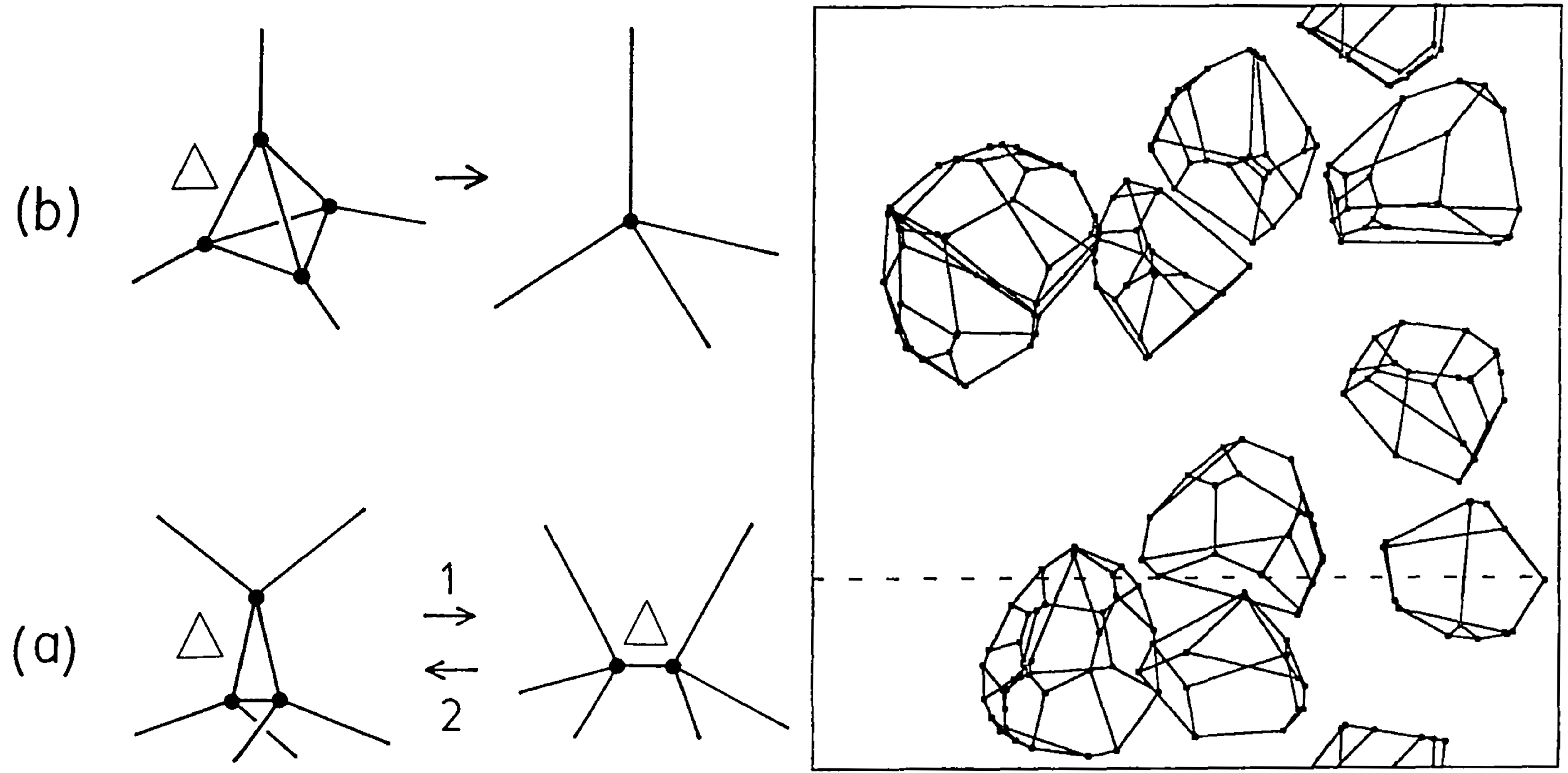}
	\caption{\label{fig:nagai1990} (left side) Three-dimensional elementary processes:(a) Recombination process, (b) Tetrahedron annihilation. (right side)  first 3D vertex simulations. Source: \cite{Nagai1990}}
\end{figure}

\begin{figure}[!ht]
	\centering
	\includegraphics[width=0.8\textwidth]{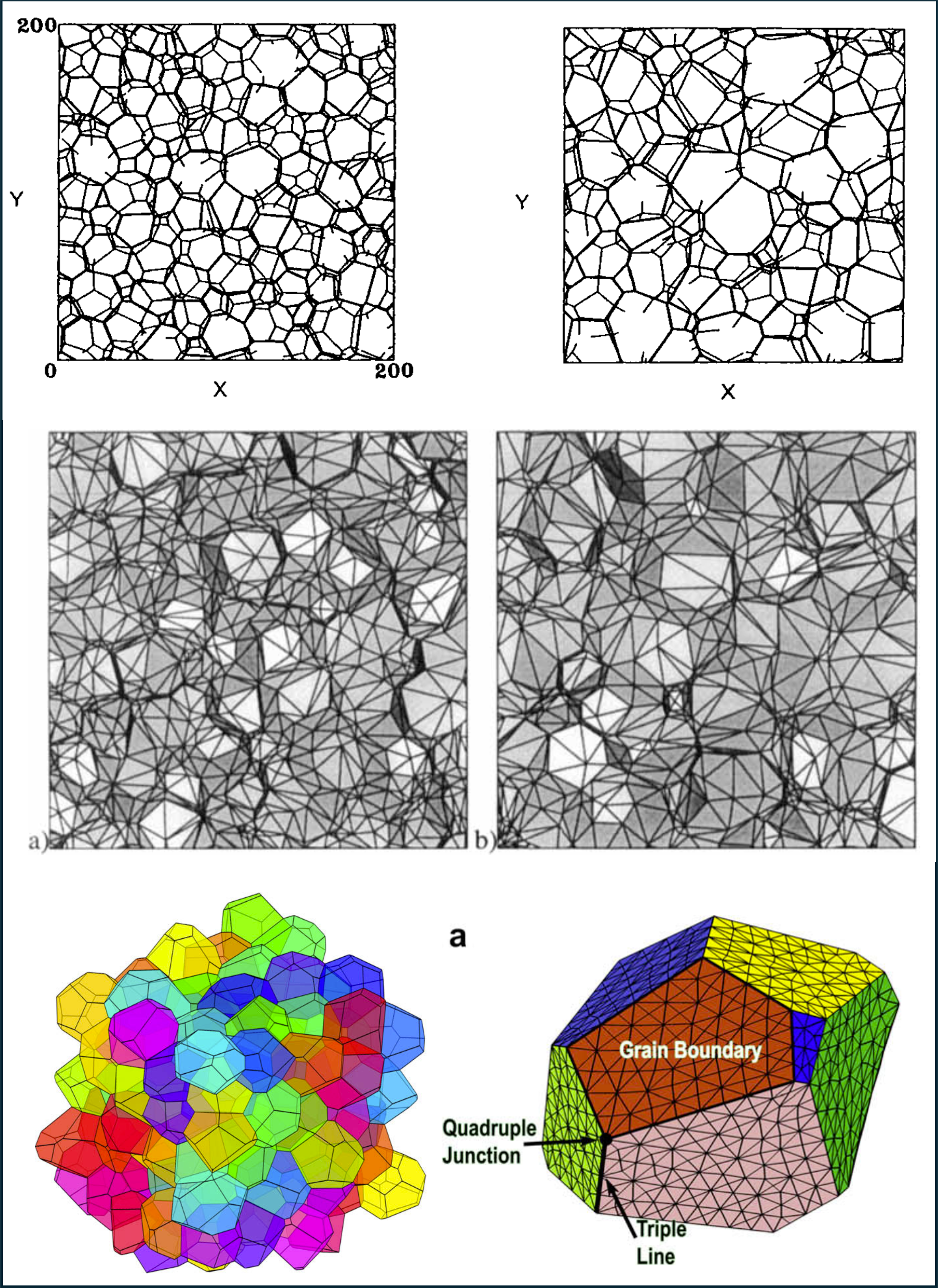}
	\caption{\label{fig:3D} First 3D representative simulations realized with the Vertex method : (top) From \cite{Fuchizaki1995}, (middle) From\cite{weygand1999a} and (bottom) From \cite{BarralesMora2008}.}
\end{figure}

The works of  Weygand et al. then lead to a series of publications in which the influence of the reduced mobility at multiple junctions and grain boundaries on grain growth was studied \cite{Weygand1998a, Weygand1999b}. However, a point quickly discussed in the literature was the ability, compared to large-scale Potts model simulations \cite{Anderson1984}, to extend the existing 2D vertex formalism to 3D while retaining the appeal of the approach, that is, its accuracy and reasonable computational cost. The first substantial work on the subject can be attributed to Nagai et al. \cite{Nagai1990}, who, starting from the Kawasaki model applied to 2D surfacic triangulation, focused on extending the possible topological operations to multiple junctions as illustrated in Fig.\ref{fig:nagai1990} (left side), but producing quite limited 3D simulations as illustrated in Fig.\ref{fig:nagai1990} (right side). Based on the work of Nagai et al., first representative simulations were proposed in \cite{Fuchizaki1995} and \cite{weygand1999a}. The size, representativeness, and complexity of the calculations then increased over the years, as illustrated by the work of Barrales Mora et al. \cite{BarralesMora2008} on the subject. These works of increasing complexity are illustrated in Fig.\ref{fig:3D}.

\begin{figure}[!ht]
	\centering
	\includegraphics[width=0.95\textwidth]{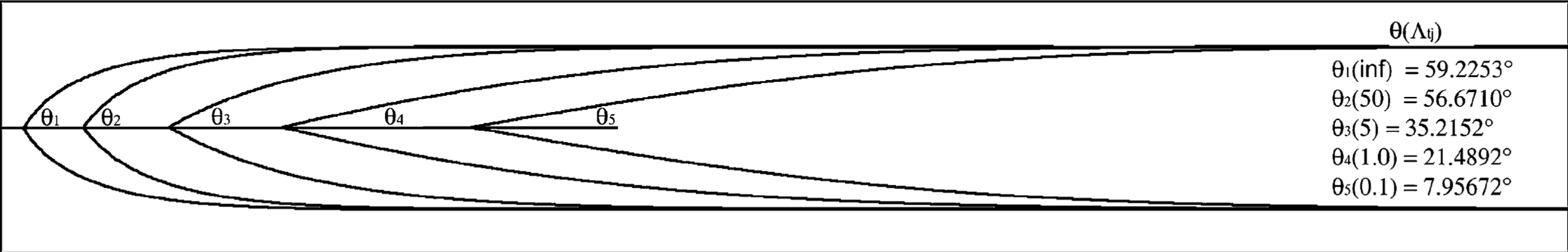}
	\caption{\label{fig:figmora} The equilibrium shape of the grain boundaries as a function of the triple junction mobility in the Barrales Mora Vertex model. From \cite{BarralesMora2010}.}
\end{figure}

Vertex models continue to be developed given their very attractive way of handling phenomenon such as grain boundary anisotropy. Indeed, one of the strengths of this approach lies in its ability to precisely describe the behavior of multiple junctions and to incorporate more or less elaborate models to describe the heterogeneity of mobility and GB energy. Typically, as illustrated by Eq.\ref{Eq:KawasakiImplicit}, even in the earlier model of Kawasaki et al. \cite{Kawasaki1989}, $\mu$ and $\gamma$ through their local GB values $\mu_{ij}$ and $\gamma_{ij}$, respectively, could be defined as dependent on the misorientation tensor. Moreover, the addition of virtual vertices also allows for consideration of the evolution of inclination along grain boundaries as an additional parameter to describe interface energy. Thus, the usual 5 degrees of freedom for describing interface energy can be immediately taken into account in these approaches. A step forward consists of being able to take into account torque terms in the grain growth driving pressure when the reduced mobility is assumed to be inclination dependent. In the context of the Vertex method, this development was first proposed by Barrales Mora as an improvement of the Weygand et al. Vertex framework \cite{BarralesMora2010}.  Fig.\ref{fig:figmora} illustrates the predictions of the Barrales Mora Vertex model \cite{BarralesMora2010} for the evolution of a triple junction with different levels of heterogeneity regarding reduced mobility.

Smith-Zener pinning mechanism by static second phase particles (SPP) was initially introduced by Smith \cite{Smith1948} and further elaborated by Zener a year later \cite{Zener1949}. Smith and Zener have highlighted how precipitates serve as barriers to grain boundaries movement, potentially stopping grain growth which presents significant industrial interest in the metal forming processes. In specific scenarios, SPP may anchor the microstructure, ultimately resulting in a characteristic limiting mean grain size (MGS) as a result of what is known as Smith-Zener pinning. This subject has thus been studied with all the existing full-field methods, and the Vertex approach is no exception to the rule. \\

\begin{figure}[!ht]
	\centering
	\includegraphics[width=0.95\textwidth]{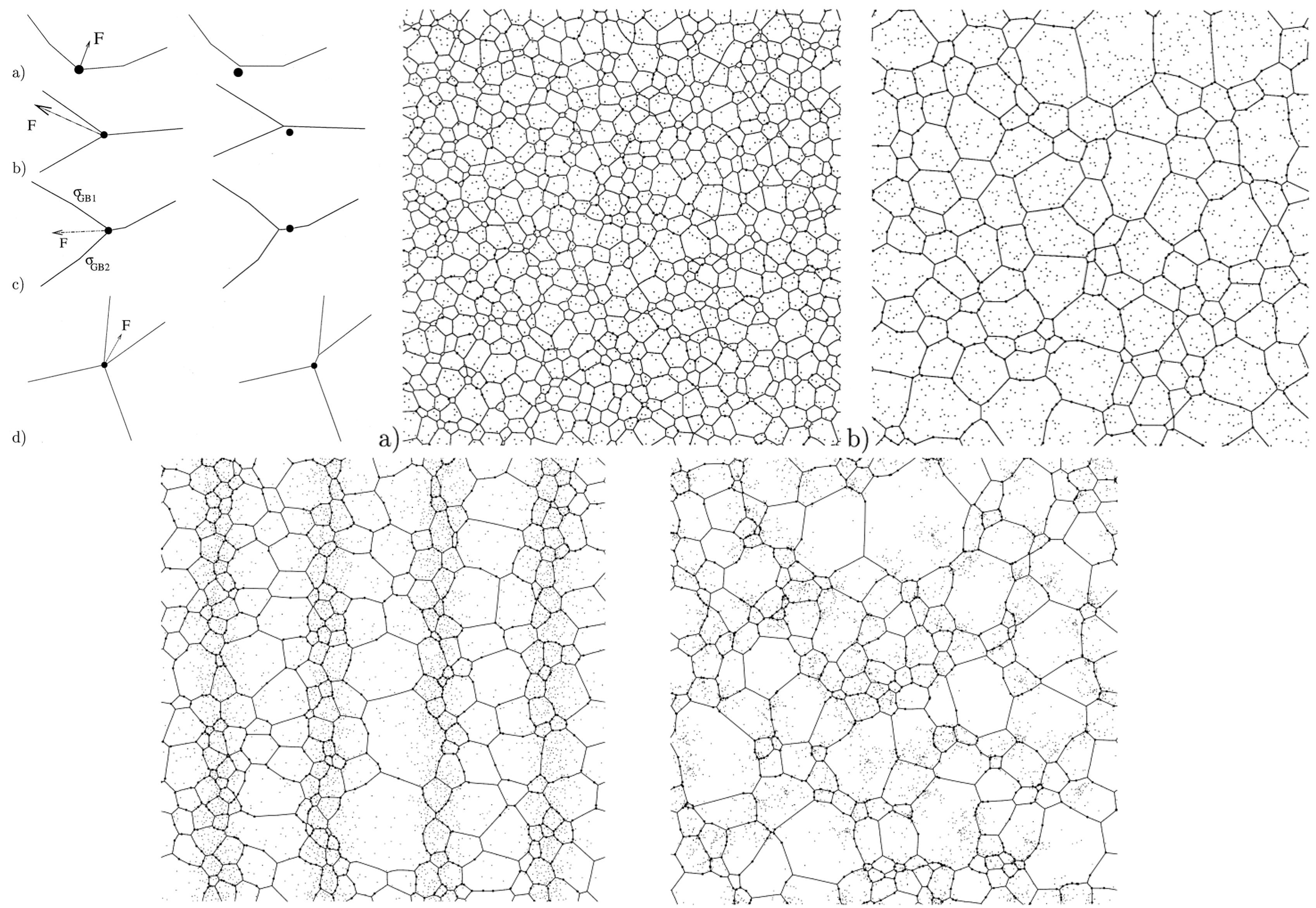}
	\caption{\label{fig:figZP1} (Top) elementary process for unpinning process as proposed by Weygand et al. and a first Vertex simulation taken into account SPP with the initial and the final stable microstructure (from \cite{Weygand1999}); (bottom) Vertex GG simulation with an heterogeneous population of SPP proposed by Lépinoux et al and based in the weygand et al. model \cite{Weygand1999} (from \cite{Lepinoux2010}).}
\end{figure}

Based on its own Vertex model \cite{Weygand1998First}, the first Vertex attempt was proposed by Weygand et al. in 1998 by considering the SPP as static vertices and a pinning position for the evolving GB \cite{Weygand1999}. An unpinning force was proposed in agreement with the Smith-Zener model in order to model unpinning events.  Fig.\ref{fig:figZP1} (top left side) illustrates the elementary process for unpinning GBs from ponctual SPP proposed in this model, whereas Fig.\ref{fig:figZP1} (top middle and right side) illustrates a 2D grain growth case with SPP leading to a static microstructure. Another simulations with the same model and heterogeneous SPP populations were proposed in \cite{Lepinoux2010} and are illustrated in Fig.\ref{fig:figZP1} (bottom).\\

The basic concept of Vertex approaches is to gain a spatial resolution dimension by focusing only on the surface of grain boundaries. Thus, this method is particularly suited to the problem of grain growth, but is much less suited to other metallurgical mechanisms involving intragranular features. As such, this method has rarely been extended to other mechanisms, such as recrystallization or solid-state phase transformations. Some exceptions can be noted and are detailed below.\\

As first tentative, we can cite the work of Weygand et al. \cite{Weygand2000} where post-dynamic recrystallization through a bulging mechanism is investigated with the introduction of a heterogeneous value of reduced mobility, but where, finally, driving pressure due to stored plastic energy is not discussed nor considered. 

\begin{figure}[!ht]
	\centering
	\includegraphics[width=0.95\textwidth]{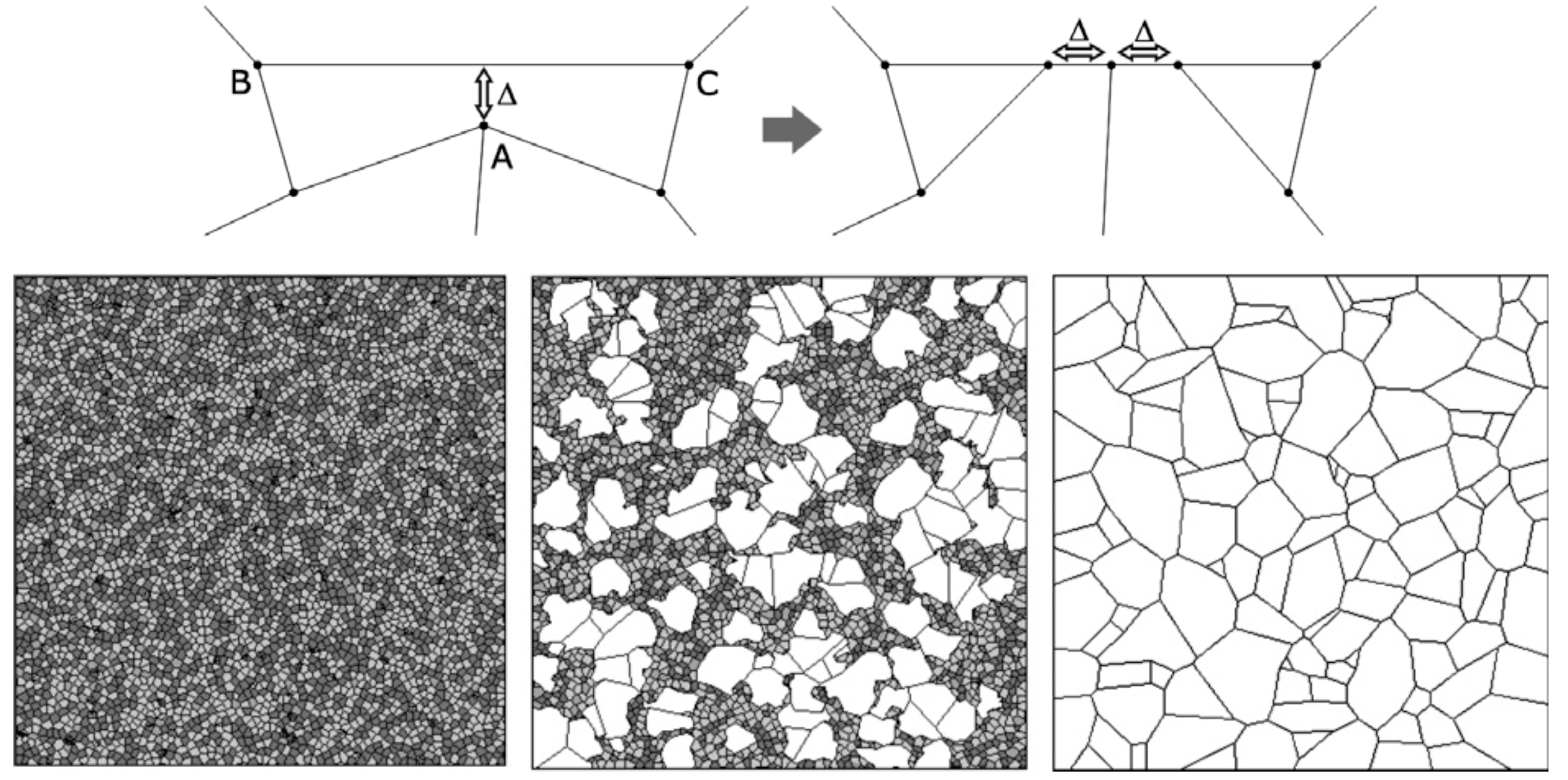}
	\caption{\label{fig:rex} (Top) New TA topological transformation proposed by Piękos et al.; (bottom from left to right) Evolution of the microstructure predicted by the Piękos et al. model: (a) initial microstructure made of 4000 grains, (b) after 50\% of recystallization, and (c) fully recrystallized. The color code depicts the stored energy. From \cite{Piekos2008b}.}
\end{figure}

A more physical approach was proposed by Piękos et al. a few years later in the context of post-dynamic recrystallization modeling \cite{Piekos2008a,Piekos2008b}. This approach is in fact more a mixture of a pure Vertex approach (without virtual vertices) for the GB description and a Monte Carlo one for the modeling of vertices coordinates evolution. The plastic stored energy is assumed to be constant per grain, and an additional topological transformation (TA) must be taken into account due to the additional driving pressure potentially leading to the evolution of a grain boundary in the direction opposite to its curvature center. This transformation is illustrated in Fig.\ref{fig:rex} (top) and a recrystallization case is shown in Fig.\ref{fig:rex} (bottom). 

One can cited, as advanced work, the model proposed in \cite{Mellbin2015}  where crystal plasticity model is coupled with a graph-based Vertex algorithm.  The crystal plasticity model is utilized within a classical finite element framework, enabling to evaluate the plastic stored energy within the polycrystal microstructure while simultaneously allowing the crystal lattices within the grains to reorient. This process impacts the advancement of recrystallization because nucleation takes place at locations where enough plastic stored energy is present, considering that the mobility and energy of grain boundaries can change on the basis of the crystallographic misorientation along the boundaries. The introduced graph-based Vertex model enables to model the topological transformations that occur in the grain microstructure and maintains awareness of the connectivity between grains. 

Finally, we can also mention the versatile and collaborative Elle Numerical Simulation Platform initiated by Jessel et al. \cite{Jessel2001}, primarily used today in geology to model deformation mechanisms, recrystallization, and grain growth in minerals. It is interesting to note that the original formalism proposed in Elle to describe the migration of grain boundaries is very close to a Vertex approach in the sense of Frost. To the author knowledge, post-dynamic recrystallization has never been investigated in 3D by the Vertex approach, nor has continuous or discontinuous dynamic recrystallization.\\

Further comments are necessary upon reviewing the state-of-the-art. First, surprisingly, it seems that Vertex approaches have not been or have rarely been confronted with the modeling of surface mechanisms involving derivatives of the curvature such as in surface diffusion mechanism (e.g in sintering or globularization). Second, the aspects of parallelization and parallel computing seem to have been little studied in the Vertex context. This can be understood in 2D given the method's efficiency for polycrystals of reasonable size, but it seems more surprising in the quest for massive 2D or 3D simulations. Third, if, as detailed previously, the issues of Smith-Zener pinning have been addressed in a Vertex context, it remains that the proposed approaches suffer from numerous limitations in terms of representativeness. Indeed, from the way to introduce SPP by static vertices, the SPP/GB interactions are poorly described, and the assumptions made seem realistic only for particle sizes that are negligible compared to the grain size, which is not representative of numerous metallic alloys. Finally, it seems that the approach has also not been adapted for the description of more realistic experimental microstructures considering the morphology of regular grain boundaries or special grain boundaries like coherent twins. All these comments have probably contributed to the development of front-tracking methods, sometimes incorporating some ideas from Vertex in managing multiple junctions but diverging in the description of grain boundaries and the modeling of their kinetics. These methods are introduced in the following section.

\section[Other front-tracking methods]{Other front-tracking frameworks for the modeling of microstructure evolution}

Front-tracking approaches different from the Vertex model can be found in the literature. In \cite{Brakke1992}, the authors introduce, in the context of 2D GG, a front-tracking methodology where instead of using a dissipative equation of motion, a variational approximation of the curvature of the interface is preferred to compute the velocity at every vertex except for those being triple junctions (see section 6.4 of \cite{Brakke1992} for more information). In fact, triple junctions are placed to respect the equilibrium in a similar manner to the works of Frost et al. \cite{Frost1988}. Moreover, the study presented in \cite{Brakke1992} is dedicated not only to the modeling of 2D GG but also to the more general context of surface energy minimization, for which the author gives many examples in 2D and 3D through a dedicated code called "SurfaceEvolver".\\

Kuprat proposed in 1998 a new code "GRAIN3D" to simulate grain growth on a volumetric mesh through the gradient weighted moving finite element (GWFE) method \cite{Kuprat2000}. However, the method's approach to topological transitions might not align with physical realities, potentially leading to significant deviations in microstructural development. On a different note, Shya and Weygand developed a technique called "generalized vertex dynamics model" for modeling grain growth on a surface mesh while managing topological changes by breaking them down into simple steps \cite{Syha2009}.\\

\begin{figure}[!ht]
	\centering
	\includegraphics[width=0.95\textwidth]{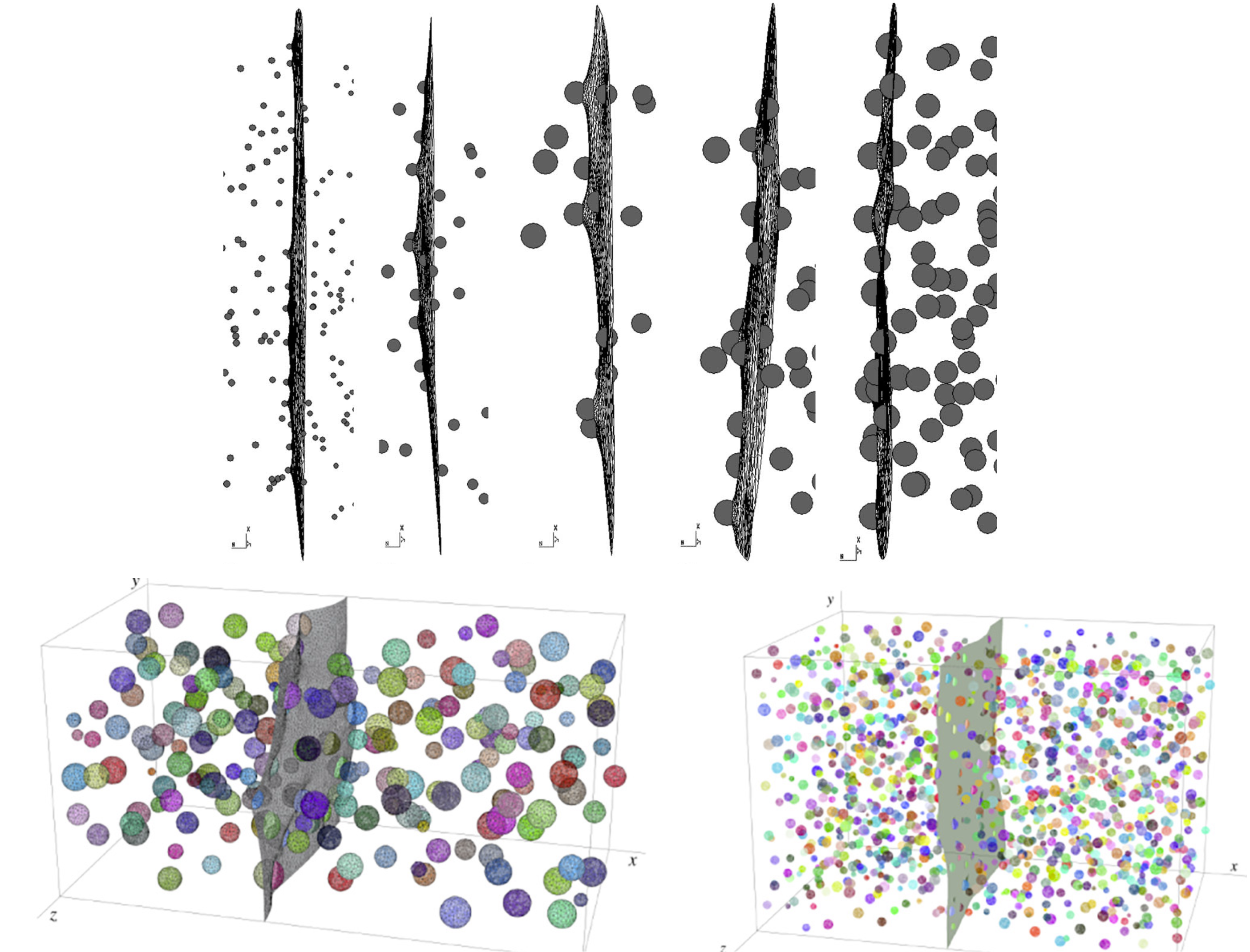}
	\caption{\label{fig:figZPCouturier} (Top) time evolution in 3D of one grain boundary evolving in a cloud of spherical static particles (from \cite{Couturier2005}), and (bottom) Similar simulation from \cite{Mohles2020}.}
\end{figure}

Another example of front-tracking models is given in \cite{Couturier2003}, where a finite element model, based on a variational formulation for grain boundary motion by viscous drag, is used to solve the equations governing grain boundary motion of an arbitrary-shaped surface and its interaction with SPP in 3D. The model was later extended in \cite{Couturier2004, Couturier2005, Couturier2003} to take into account motion by curvature flow but in the context of a single grain boundary and led to impressive simulations of Smith-Zener pinning mechanisms comparatively to the state of the art, as illustrated in Fig.\ref{fig:figZPCouturier}(top). However, real polycrystal structures were not tested with this approach.  A similar discussion was recently proposed in \cite{Mohles2020} with comparable simulations, but a different approach to describe the interaction between SPP and grain boundary allowing to consider a richer description of SPP in terms of shape and energy relationship with the matrix as illustrated in Fig.\ref{fig:figZPCouturier}(bottom).\\

Another Front-Tracking model was proposed in \cite{Becker2008} where the movement of grain boundaries is defined through the minimization of the total energy of the system by moving each node along the energy gradient towards a lower total energy state. In this model, the energy term can be defined via the plastic stored energy, the grain boundary energy, and the non-conservative reaction terms. The evaluation of the energy gradient field is determined for every node via a local finite-difference method centered at the node in question, and the movement is made in the direction of the maximal energy reduction. The authors tested this model in the context of isotropic and anisotropic grain boundary migration, where melting was simulated through a second phase, attributing an "energy-based" melting-fraction function, where the more the melting fraction deviates from the target value, the more the melting/crystallization energy dominates. However, no information regarding the grain boundary anisotropy function was given, even though some results show anisotropic behavior between the solid-solid and solid-liquid boundaries.\\

\begin{figure}[!ht]
	\centering
	\includegraphics[width=0.95\textwidth]{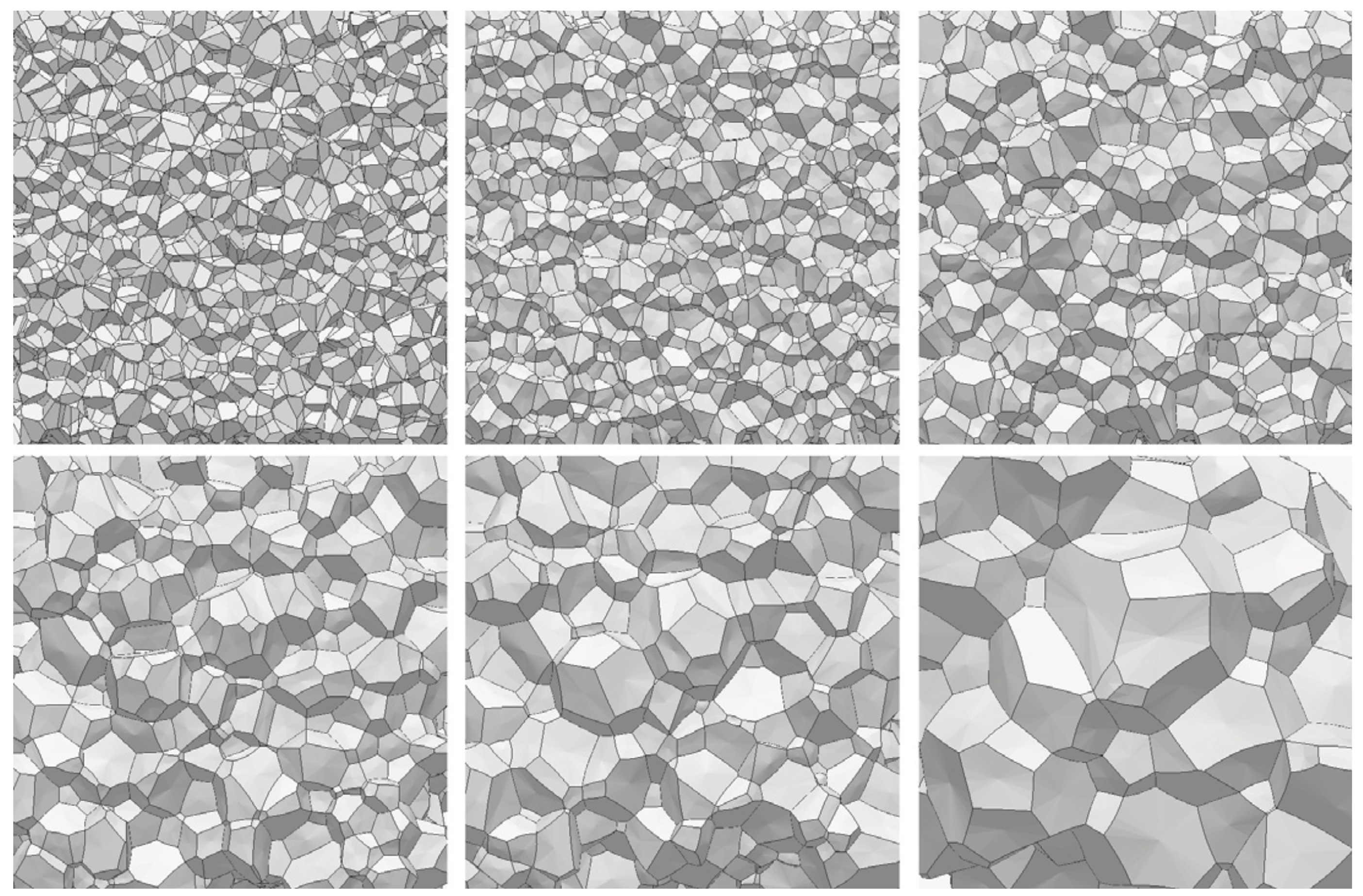}
	\caption{\label{fig:figLazar} From top to bottom and left to right: time evolution of a 3D pure grain growth case realized with a  MacPherson-Srolovitz based front-tracking method. From \cite{Lazar2011}.}
\end{figure}

In 2011, Lazar and colleagues introduced a method based on a discretized version of the MacPherson-Srolovitz relationship \cite{macpherson2007neumann} to simulate ideal grain growth on a surface mesh and with uniform grain boundary characteristics, necessitating only a limited number of topological changes \cite{Lazar2011}. Fig.\ref{fig:figLazar} illustrates a 3D ideal grain growth simulation performed with this method.\\

Recently, Eren et al. \cite{Eren2021,Eren2022} proposed a new 3D approach based on 3D finite element mesh. This approach presents the interest to promote optimal fitted mesh in terms of numerical cost to model grain growth \cite{Eren2022}. Moreover, a large investigation of 3D topological transitions was performed in the context of the use of unstructured finite element mesh \cite{Eren2021}. This choice also enables to use efficient and parallelized finite-element libraries for FE resolution and remeshing operations. Today, it is the most advanced work in the context of 3D mesh-based front-tracking approaches for grain growth.\\

\begin{figure}[!ht]
	\centering
	\includegraphics[width=0.99\textwidth]{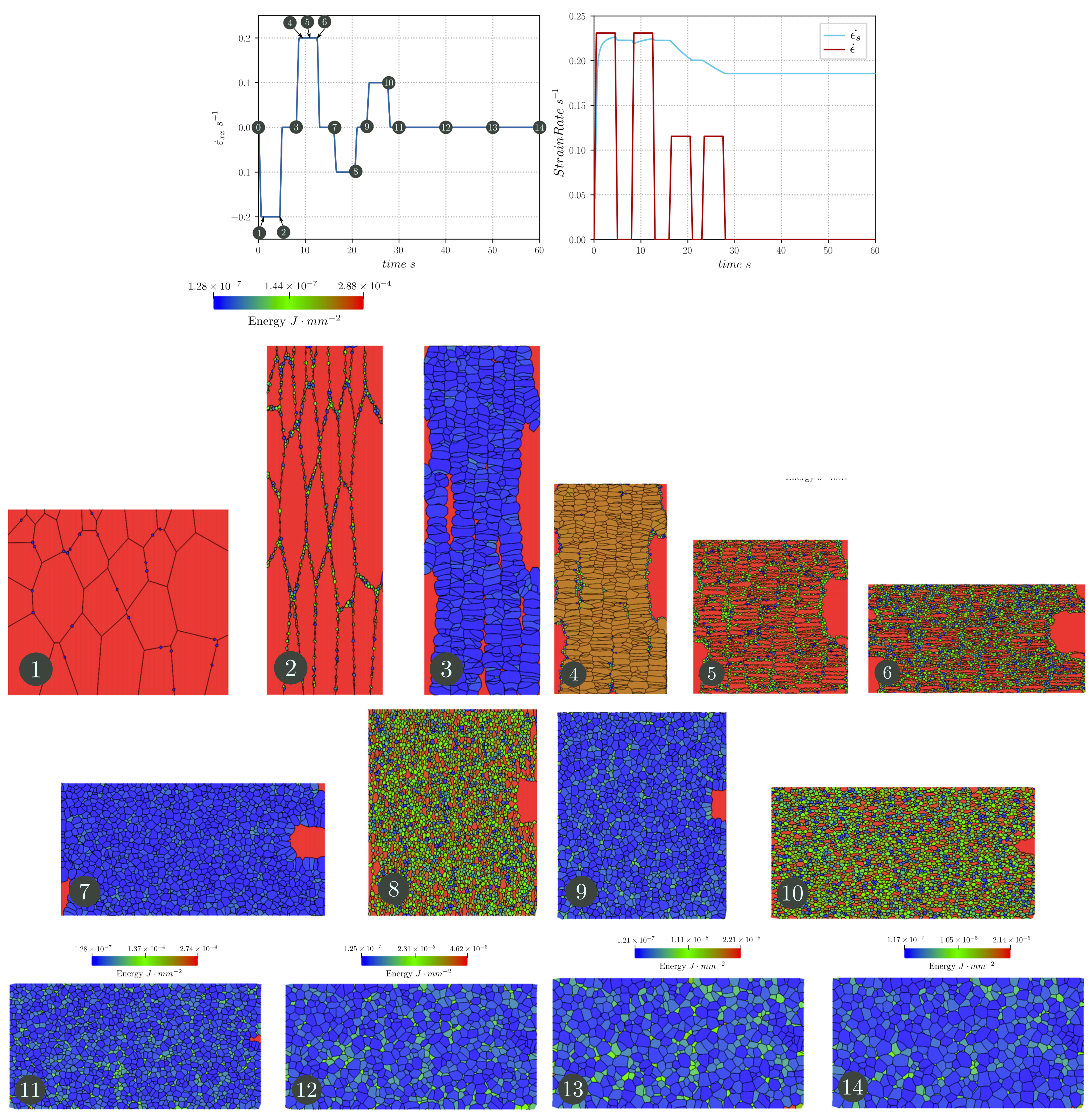}
	\caption{\label{fig:figTRM} This figure illustrates the use of TRM code in the context of the modeling of a complex industrial thermomechanical path applied to a 304L stainless steel and described at the top (strain and strain rate, the temperature being constant and equal to \SI{1080}{\celsius}). The predicted microstructures at each numerated position in the path are described by exhibiting the stored energy field and the grain boundaries in black. This simulation involves discontinuous dynamic recrystallization, meta-dynamic recrystallization and grain growth mechanisms. From \cite{Florez2021a}.}
\end{figure}

In the context of 2D generic front-tracking models for numerous mechanisms, one can also cite the recent developments of Florez et al. concerning the ToRealMotion (TRM) code for "topological remeshing in Lagrangian framework for large interface motion" \cite{Florez2020}. The TRM approach introduces the concept of using unstructured FE meshes for the detailed representation of grain interiors. This strategy is adopted for several key reasons: first, to incorporate intragranular data such as stored energy to model recrystallization; second, to accurately simulate significant domain deformations and discontinuous dynamic recrystallization mechanisms \cite{Florez2021a}; and third, to enhance the parallel processing capabilities of the method, surpassing those of conventional front-tracking models \cite{Florez2021b}. This method was also improved, considering the treatment of multiple junctions proposed by Barrales Mora \cite{BarralesMora2010}, to deal with anisotropic reduced mobility \cite{Florez2021c} and torque terms \cite{Florez2022}. Recently, this formalism has also been extended to account for static or dynamic second-phase particles, by discretizing those of large sizes and using pinning points for those of smaller sizes \cite{Florez2025}. Fig.\ref{fig:figTRM} illustrates the use of TRM in the context of discontinuous dynamic recrystallization modeling with a complex industrial thermomechanical treatment that includes large deformation and involves discontinuous dynamic recrystallization, meta-dynamic recrystallization and grain growth mechanisms.

\section{Conclusion}
In conclusion, Vertex approaches represent a critical methodology in the field of computational materials science. These methods, which take advantage of the precise discretization of grain boundaries, offer a sophisticated framework for understanding and predicting the evolution of microstructures under various conditions. Vertex models stand out for their ability to accurately model complex grain boundary movements, including curvature-driven growth with anisotropic grain boundary properties. Furthermore, their computational efficiency makes them particularly suitable for study of large polycrystals compared to front-capturing methods such as multi-phase field and level-set ones. Due to some challenges such as the need for more generic models to capture all relevant phenomena and the difficulty to develop 3D models, Vertex approaches continue to evolve mainly through new front-tracking approaches where some ingredients of the original Vertex approaches are used such as the treatment of the multiple junctions. With ongoing developments in computational power and numerical techniques, these methods are set to provide even deeper insights into grain boundary modeling, supporting advancements in materials design and engineering.

\bibliography{main} 

\end{document}